# Observation of 2D Cherenkov radiation

Yuval Adiv[1*], Hao Hu[2*], Shai Tsesses[1*], Raphael Dahan[1], Kangpeng Wang[1], Yaniv Kurman[1], Alexey Gorlach[1], Hongsheng Chen[3], Xiao Lin[3], Guy Bartal[1] and Ido Kaminer[1]

[1]Andrew and Erna Viterbi Department of Electrical Engineering, Technion, Israel Institute of Technology, 32000 Haifa, Israel

[2]School of Electrical and Electronic Engineering, Nanyang Technological University, Nanyang Avenue, Singapore 639798, Singapore

[3]Interdisciplinary Center for Quantum Information, State Key Laboratory of Modern Optical Instrumentation, ZJU-Hangzhou Global Science and Technology Innovation Center, College of Information Science and Electronic Engineering, Zhejiang University, Hangzhou 310027, China

**Abstract:**

**For over 80 years of research, the conventional description of free-electron radiation phenomena, such as Cherenkov radiation, has remained unchanged: classical three-dimensional electromagnetic waves. Interestingly, in reduced dimensionality, the properties of free-electron radiation are predicted to fundamentally change. Here, we present the first observation of Cherenkov surface waves, wherein free electrons emit narrow-bandwidth photonic quasiparticles propagating in two-dimensions. The low dimensionality and narrow bandwidth of the effect enable to identify quantized emission events through electron energy loss spectroscopy. Our results support the recent theoretical prediction that free electrons do not always emit classical light and can instead become entangled with the photons they emit. The two-dimensional Cherenkov interaction achieves quantum coupling strengths over two orders of magnitude larger than ever reported, reaching the single-electron–single-photon interaction regime for the first time with free electrons. Our findings pave the way to previously unexplored phenomena in free-electron quantum optics, facilitating bright, free-electron-based quantum emitters of heralded Fock states.**

**Introduction**

Interactions between free electrons and light are of prime importance for fundamental science, applications, and future technology. Examples include Compton scattering, which is utilized in radiation therapy and spectroscopy [1]; photon-induced nearfield electron microscopy (PINEM), which exposes femtosecond physical phenomena at the nanoscale [2–4]; dielectric laser accelerators, which enable chip-scale particle acceleration schemes [5–7]; and cathodoluminescence, which provides powerful microscopy capabilities and facilitates novel nanophotonic light sources [8–12].

Cherenkov radiation (CR) is a well-known effect in this family of interactions, first discovered in 1934 [13,14] from charged particle surpassing the phase velocity of light in a medium and emitting an electromagnetic shockwave (often seen as a bluish glow). This discovery was the first instance of a wider concept of spontaneous emission as a result of a phase-matching between the free electron and the emitted light. This general concept requires the particle velocity to match the phase velocity of light along the particle's propagation direction, defining a characteristic emission angle as a function of the particle velocity, when greater than the phase velocity of light.

Today, the term Cherenkov effect is widely used to describe a variety of phenomena that arise from phase-matching between free charged particles and photons in a myriad of materials. CR-type effects have been proposed and observed in a wealth of artificially engineered materials in which the phase velocity of light can be flexibly tailored [15,16], e.g., photonic crystals [17,18], hyperbolic media [19], gain media [20], and negative-index metamaterials [21–23]. Types of CR were also examined for charged particles traveling in close proximity to a material [24,25]. While research in CR led to a rapid succession of theoretical and experimental discoveries spawning many applications [15,26–30], it is still usually observed as a classical wave phenomenon occurring in 3D geometries.

Interestingly, free-electron–light interactions change dramatically with dimensionality, as was extensively explored theoretically [31–35]. While the transverse nature of 3D propagating waves restricts the spectral and angular emission of conventional CR (3D-CR) to a broadband radiation cone, CR emitted into waves that are forced to propagate in two-dimensions (2D-CR) was predicted to exhibit an intense narrow spectrum propagating primarily parallel to the electron's trajectory [9,34–39] (Fig. 1). The effect of 2D-CR was initially predicted in the form of surface plasmon polariton (SPP) modes, also known as Cherenkov-Landau surface shockwaves [36,37]. The key to this prediction was the *propagating nature* of SPPs, enabling them to maintain phase-matching with the charged particle when it moves parallel to the surface with velocity surpassing the SPP phase velocity. In other cases, SPPs emitted via the phase-matched coupling can in turn couple out to free space photons and become 3D-CR, as predicted in refs [38,39]. In any dimensionality, the radiation angle $\theta$ (relative to the particle's trajectory) satisfies $\cos\theta = v_p(\omega)/v_e$, where $v_e$ is the charged particle velocity and $v_p(\omega)$ is the phase velocity of the wave, that depends on the frequency $\omega$.

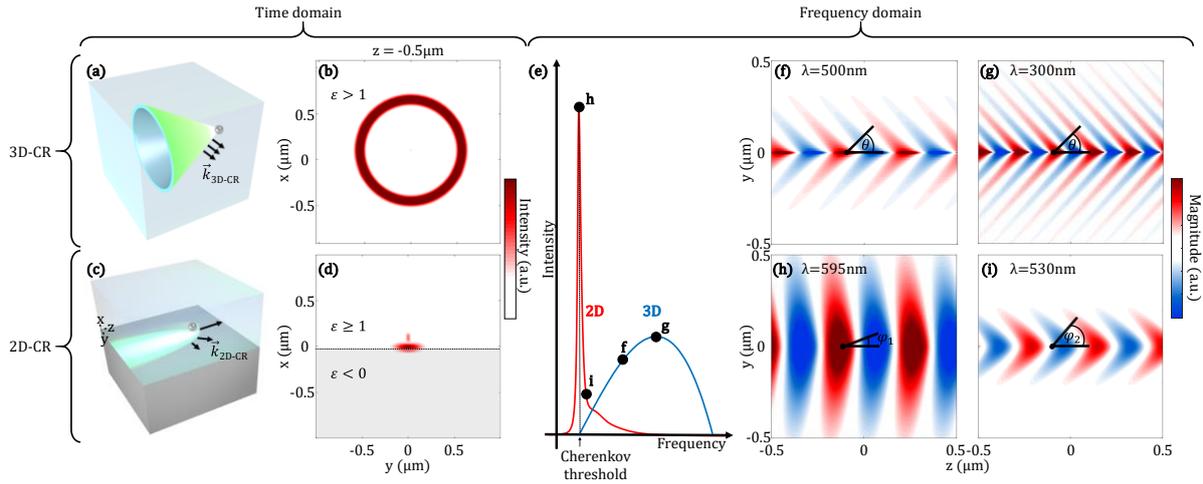

**Fig. 1 | 3D vs. 2D Cherenkov radiation (CR): Emphasizing their fundamentally different features. (a,b)** Illustration of 3D-CR. The radiation is emitted in a set of wavevectors ($\vec{k}_{3D-CR}$) which form a cone (a ring cross section) around the direction of electron velocity, with zero intensity at the direction of motion. **(c,d)** Illustration of 2D-CR. The radiation is emitted in a set of wavevectors ($\vec{k}_{2D-CR}$) along the 2D surface, with peak intensity at the electron's direction of motion (shown in the cross section by the peak intensity near the surface). **(e)** Schematic comparison of 3D-CR and 2D-CR spectra. Since the dispersion of conventional materials is relatively small and isotropic, the 3D-CR spectrum is broad and peaks at a frequency higher than the Cherenkov threshold. In contrast, the 2D-CR spectrum is narrow, peaks at the Cherenkov threshold, and decays sharply for large frequencies. **(f)–(i)** Comparison between the electric field profiles in 3D-CR and 2D-CR for selected wavelengths brought as examples of the frequency dependence (corresponding to the points in panel (e)). These panels highlight the dispersive emission angle of 2D-CR versus the non-dispersive emission angle of 3D-CR. $\theta$ (in panels f,d) and $\varphi_1$, $\varphi_2$ (in panels h,f) are the angles between the electron trajectory and the direction of electromagnetic wave emission, $\vec{k}_{3D-CR}$ and $\vec{k}_{2D-CR}$, respectively.

The unique features of 2D-CR make it a promising platform to achieve a versatile, tunable, and ultrafast conversion mechanism from electrical signal to plasmonic excitations [34]. Recent studies observed analogues of 2D-CR in metasurfaces and other nanophotonic systems using an electromagnetic wave that replaced the emitting charged particle by a mathematical analogue [40,41]. However, no experiment thus far has ever reported the observation of 2D-CR by free electrons or by any charged particle. Therefore, none of the unique features emanating from the 2D nature of the emission has been observed.

Here we present the first observation of 2D-CR by free electrons. Utilizing a dispersion-engineered nano-photonic structure, we match the velocities of the electrons to that of the 2D waves guided in the structure to fulfill the required phase-matching. This way, we achieve record-high emission rates and the unique spectral features associated with 2D-CR. We utilize this enhanced interaction to provide an experimental illustration of a recent paradigm shift, predicting the entanglement of the electrons with the wave they emit.

The radiation in every optical environment that is created by free or bound electrons, including *all* CR effects, can be conveyed in the language of *photonic quasiparticles* (PQPs) [42]. The emitted PQP can be either a photon in a 3D dielectric medium [13,14,17,18], a polariton in a 2D material [34,43,44], a surface plasmon polariton on an interface between materials [45,46], a phonon polariton in a crystalline solid [47], etc. All these are forms of propagating PQPs, defined by a relatively long lifetime (longer than the cycle of the PQP) and long spatial extent (longer than the wavelength of the PQP). Their propagating nature enables them to satisfy phase-matching with the emitting electron, thus becoming part of a Cherenkov-type process. Other types of PQPs such as bulk plasmons and surface plasmons [48–56] are non-propagating (due to a below-cycle lifetime or spatial extent) and thus cannot take part in Cherenkov-type processes. Hereinafter, we focus on *propagating* PQPs.

## Experimental settings

Obtaining 2D-CR requires a careful design of the PQP's dispersion such that it is phase-matched with the moving electron. We utilize a metal-dielectric multilayer structure supporting hybrid photonic–plasmonic modes [57,58] whose dispersion can be carefully tuned by the system's geometry [59,60]. Hence, we design such a structure (Fig. 2 (a)–(d)) to match the phase velocity of the PQPs to the electron velocity in a transmission electron microscope (TEM) – providing the ability to achieve the Cherenkov condition over a range of PQP energies (2.083 eV to 2.295 eV) and electron kinetic energies (93 keV to 200 keV).

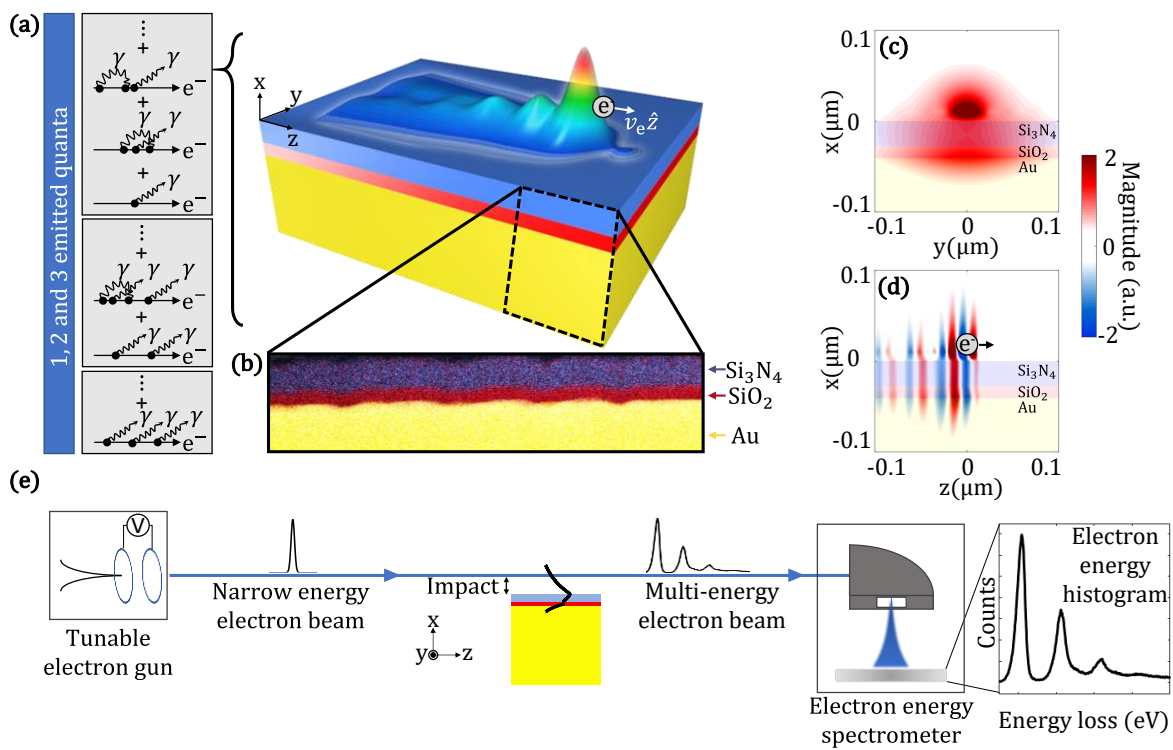

**Fig. 2 | The experimental setup for the 2D-CR measurement. (a)** Schematics of a free electron traveling above our metal-dielectric multilayer structure emitting 2D-CR. The YZ cross section of the radiated electric field (z-component) is presented, highlighting its co-propagation with the electron. The schematic represents the emission of a single quanta of PQP (exemplified by the Feynman diagrams of 1 photon emission), which is part of the joint electron–photon quantum state. Higher-order processes are also possible as shown in the other Feynman diagrams. **(b)** Cross-section image of the metal-dielectric multilayer structure. Layers are presented in different colors according to an energy-dispersive X-ray (EDX) spectroscopy measurement (see Appendix 1). **(c,d)** The field distribution of the PQP created by 2D-CR through its XY (c) and XZ (d) cross sections. To maximize the overlap with the free electron, the sample is engineered to support a confined (c) PQP with a large field distribution in its evanescent tail that extends out of the structure (d). **(e)** Illustration of the experiment, wherein an electron beam propagates parallel to the sample surface and emits multiple quanta of PQPs. The emission events are imprinted on the electron energy spectrum as discretized energy loss events measured in the electron energy loss spectrometer (EELS). Inset shows a characteristic EELS measurement, featuring three 2D-CR peaks beyond the zero-loss peak. The distance between the electron beam centroid to the sample surface is the impact parameter.

Our experimental setup, shown in Fig. 2(e), includes a highly directional electron beam in grazing angle conditions, positioned at a distance (impact parameter) of tens of nanometers from the surface and sustained along tens to hundreds of microns (see ref [61] and Appendix A for more details about achieving grazing angle interaction in our setup). This setup allows us to control several degrees-of-freedom that affect the 2D-CR. By varying the electron kinetic energy, the energy of the emitted PQP is tuned, while varying the impact parameter and interaction length (Fig. 2(e)) affects the emission probability (the coupling strength) between the electron and the PQPs. These properties are then inferred from electron energy loss spectroscopy (EELS). Energy shifts as small as ~10 meV could be identified (see Appendix A).

**Results**

Figure 3(a) presents the first EELS peak, showing that it red-shifts for higher electron velocities, as expected from the CR phase-matching theory. This red-shift has never appeared in previous EELS measurements, since all EELS experiments so far did not include extended phased-matched interactions. The EELS arising from interactions with non-propagating PQPs is mostly independent of the electron velocity (though slighter blue-shifts can occur for higher electron velocities due to a frequency dependence of the electron coupling strength to optical excitations).

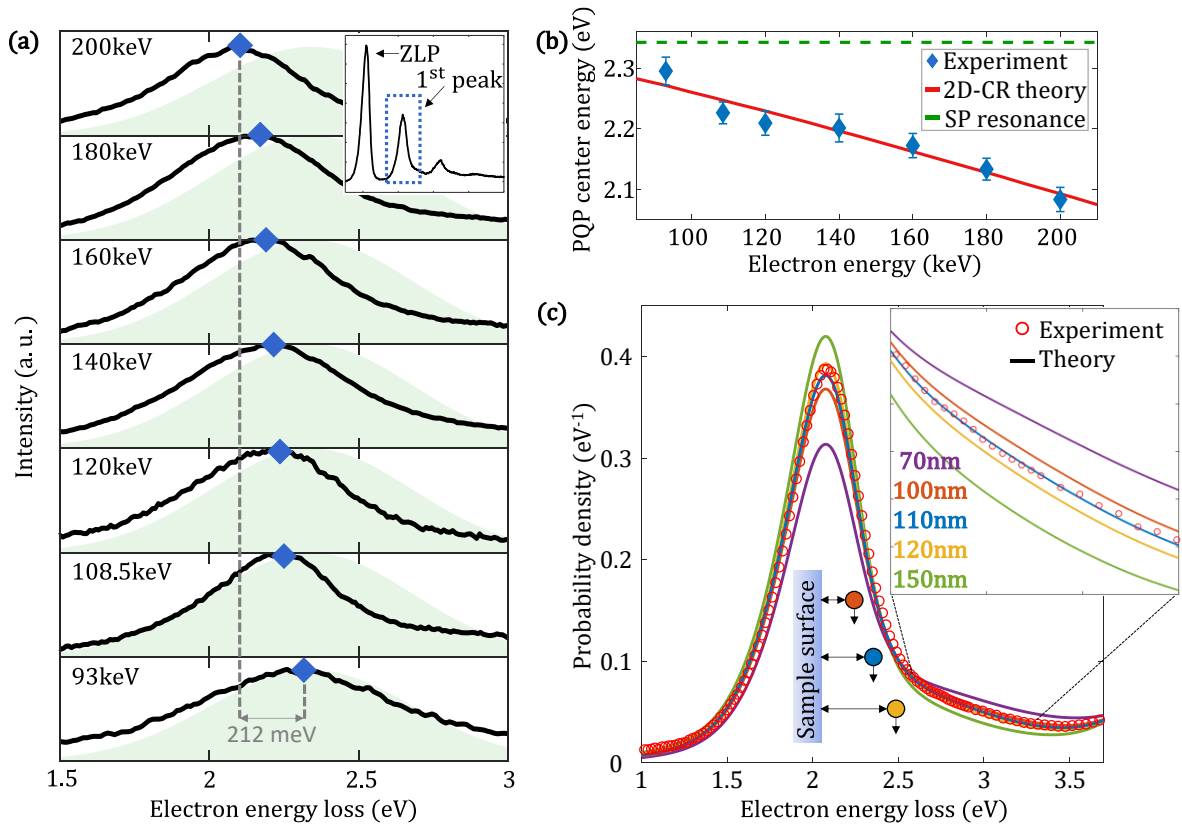

**Fig. 3 | First observation of 2D-CR: satisfying the Cherenkov phase-matching condition. (a)** The first EELS peak of each measurement (solid black line) for electrons grazing the multilayer structure. Increasing the electron energy results in a red-shift of the peak, a signature of 2D-CR. The shifting peak stands in contrast to EELS of common non-propagating surface plasmon resonances, which remain constant when changing the electron energy (light green). The inset presents a characteristic raw measurement. **(b)** Comparison of experiment and theory. The measured peak energies from (a) (blue diamonds) are compared to theory of 2D-CR into PQPs (red, see calculation in Supplementary Note 1) and to the excitation of non-propagating surface plasmon (SP) mode (dashed green line). Only the former shows agreement with our measurements. **(c)** Analysis of the EELS peak shape of 2D-CR: comparison of theory and experiment. The first EELS peak is shown for a range of impact parameters, exhibiting the hallmark asymmetrical profile of the 2D-CR emission spectrum, similar to Fig. 1(e). The spectrum has its peak at the Cherenkov threshold, a feature of 2D-CR that differs substantially from the case of 3D-CR. The good match of the simulated peak shape and location (colored solid lines) with the measured data (red circles) provides additional evidence for our observation of 2D-CR and allows extraction of the effective impact parameter of the electron beam centroid with 10 nm accuracy. The inset shows a zoom-in, which provides a more precise estimate of small changes in the impact parameter. The upward slope at the right edge of the spectrum arises from the second EELS peak.

In Fig. 3(b), we show how the 2D-CR theory predicts the locations of the first EELS peak in every measurement. Each peak emission frequency can be approximately determined by the intersection point of a line with slope equals to the electron velocity with the PQP dispersion (Appendix Figure 1).

This intersection explains the observed red-shift of the emitted PQP with increasing electron energy. Had the electron only interacted with non-propagating surface or bulk plasmons, as in previous EELS experiments that also showed multiple emission events, e.g., refs [48–56], the EELS peaks would not been altered as a function of the electron velocity and the spectral red-shift could not have been observed.

Apart from the specific spectral peak locations, the 2D-CR theory (Supplementary Note 1) also matches the spectral shape of the measured peak (Fig. 3(c)). Notably, the asymmetric profile is a hallmark of 2D-CR into lossy PQPs [34]. The excellent agreement between the theory and experiment enabled us to more precisely determine the experimental parameters since the spectral profile of 2D-CR is highly sensitive to the impact parameter with a 10 nm accuracy.

**The quantum photonic nature of 2D-CR**

The observed 2D-CR provides new insight on the *quantum* nature of free-electron radiation, which has recently been under intense theoretical investigation [62–70]. The CR effect, like any other form of free-electron radiation, was surmised to be an emission process of *classical* light [9,67–70] (i.e., a Glauber coherent state in the quantum optics nomenclature). Recent theoretical advances [62–70] created a paradigm shift in the understanding of spontaneous emission from free electrons: only an electron with a (coherently) wide energy-uncertainty could emit classical light. An electron with energy-uncertainty narrower than the energy of its emitted photons should become entangled with these photons [67–70].

Our results provide experimental evidence in support of this new paradigm, by having a narrow-enough electron energy uncertainty. Thus, each electron becomes entangled with the PQPs it emits, which enables measuring the emission properties – both spectral shape and number of emission events – from the electron energy loss. This quantized nature is found in some of our EELS measurements, exhibiting multiple loss peaks with a separation of $\hbar\omega_0$, where $\omega_0$ is the peak frequency of the PQP (Fig. 4(a,b)). Each energy loss peak represents an emission of an integer number of PQPs. For comparison, for the electron to emit classical light (or classical 2D-CR), its energy-uncertainty must have been wider, as shown in Fig. 4(c).

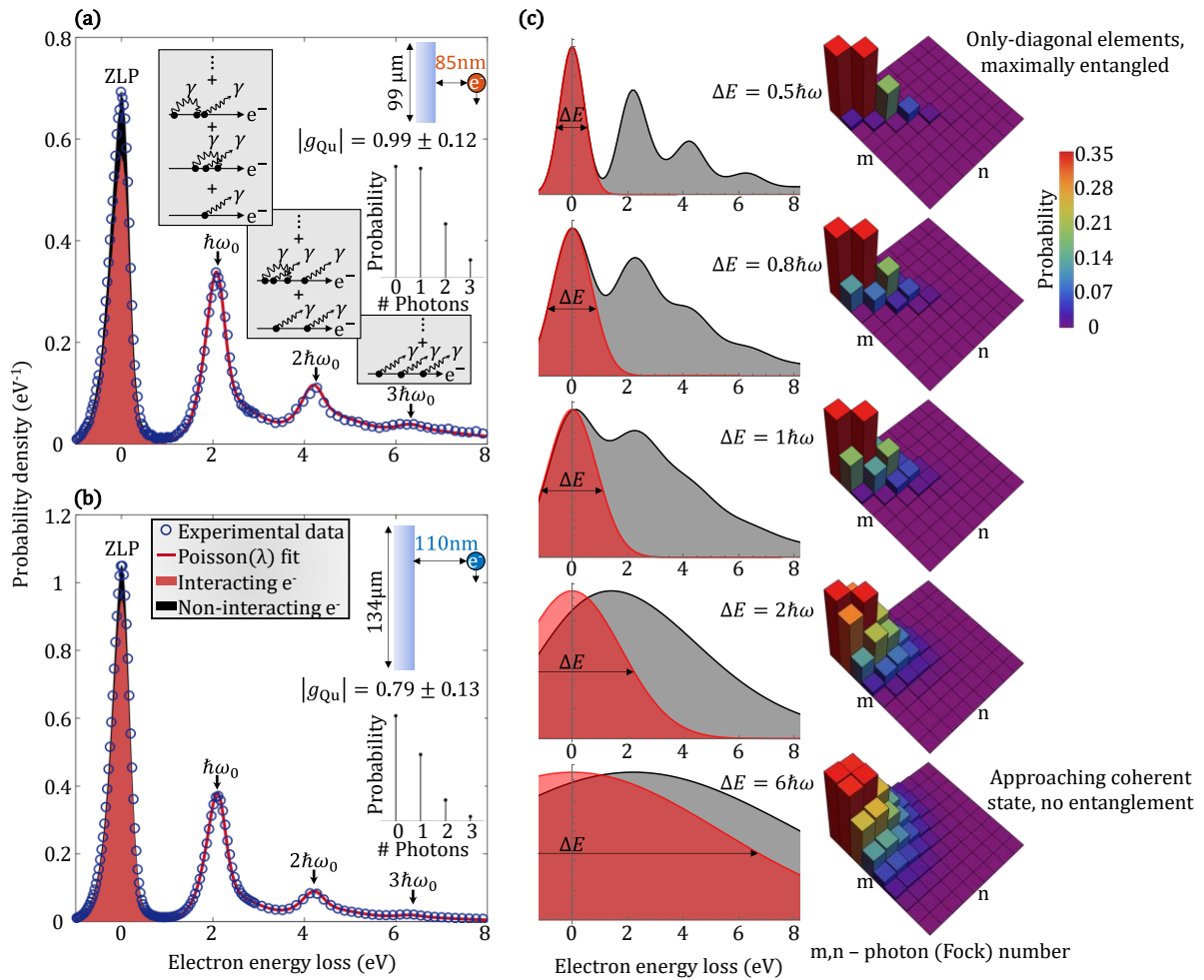

**Fig. 4 | Measurement of quantized radiation by free electrons: high-order emission of multiple photonic quasiparticles.** **(a,b)** EELS measurements of 200 keV electrons (blue circles) for two different impact parameters and effective interaction lengths reveal the quantized photonic nature of 2D-CR by showing multiple peaks equally separated by the PQP energy. The suggested model (solid red curve, Eq. 2) follows the recorded data precisely and allows the extraction of various parameters: (1) the total probability to emit zero, one, two, or three PQP (showed in the inset). This probability follows a Poisson distribution according to the Feynman diagrams in panel (a) that include multiple emission and reabsorption events in each multi-photon emission process; (2) a fairly substantial quantum coupling strength $|g_{Qu}| = \sqrt{\lambda}$ with $\lambda$ being the Poisson distribution parameter; (3) the ratio between the electrons that did and did not interact (red and black filling of the ZLP peak, respectively). **(c)** Theoretical analysis of the joint electron-PQP state, showing conditions for the emitted photons to be entangled with the emitting electron, and how they depend on the electron (coherent) energy uncertainty. EELS (left column) and the corresponding photonic density matrix in the Fock basis (for the peak frequency; right column) were simulated for different values of electron energy uncertainty using $|g_{Qu}|^2 = 1$, to resemble the case in panel (a). Each EELS is found by tracing-out the photonic part of the joint density matrix, whereas each photonic density matrix is found by tracing-out the electron part. The top row shows the case of an electron energy uncertainty narrower than the photon energy that creates a maximally entangled electron-PQP state. The bottom row shows the case of an electron energy uncertainty much wider than the photon energy, which can represent a short duration electron pulse that creates a separable electron-PQP state with a Glauber coherent (classical) PQP state. See Supplementary Information and Appendix 1 for additional details regarding the quantum description of the process, parameter extraction, and data analysis.

Figure 4(c) explains the transition from classical to entangled free-electron radiation when reducing the electron's energy uncertainty. The bottom row in Fig. 4(c) shows that the photonic density matrix of light emitted by an electron with a wide energy uncertainty can approach a Glauber coherent state, i.e., classical light. The emission of classical light further requires the electron temporal duration to be shorter than the emitted field cycle (which given the wide energy uncertainty, is possible but not necessary). This condition is commonly occurring for radiation in the radiofrequency and microwave regimes, where the field cycle is long. However, this logic cannot be easily extended to the optical regime. Indeed, this condition is violated in our experiment, and in fact, we expect it to be similarly violated in almost all electron radiation experiments in the optical regime (and in higher frequencies such as X-rays).

The measured EELS (Fig. 4(a,b)) closely resemble the top row in panel (c). There, the photonic density matrix of light contains near-zero off-diagonal terms, implying that the photons were nearly fully entangled with the electron. Therefore, the EELS measurement provides indirect evidence showing that the CR process causes electrons to become entangled with the PQPs they emit (as was predicted very recently in ref [69]). It is important to note that the theory supporting our analysis, has been derived by a few recent studies that created a paradigm shift of how we understand the quantum nature of free-electron–photon interactions [67–70]. Our measurement is the first to provide indirect experimental evidence for this new understanding by using electron emission into propagating modes. Direct evidence for the underlying entanglement would require a coincidence measurement between the electron and photons.

**Theoretical description of the 2D-CR quantum photonic nature**

The EELS measurements presented in Fig. 4(a,b) allow for resolving individual emission events of single PQPs. Such results cannot be reproduced by classical electromagnetism since it ignores the photonic nature of light. Thus, to quantify the efficiency of PQP emission, we recall that each CR process can be described as spontaneous emission by an electron into photonic vacuum fluctuations that are phased-matched with the electron [42,71]. The quantized nature of the PQP emission can be captured by a compact scattering matrix description [72]:

$$\hat{S} \triangleq \exp[g_{\mathrm{Qu}} b a^\dagger - g_{\mathrm{Qu}}^* b^\dagger a] \tag{1}$$

where $b$ $(b^\dagger)$ is the electron energy ladder operator that decreases (increases) the electron energy by a discrete amount, and $a$ $(a^\dagger)$ is the annihilation (creation) operator of a PQP; $g_{\mathrm{Qu}}$ is the quantum coupling strength, which is proportional to $\sqrt{L_{\mathrm{eff}}} e^{-x_0|k_\mathrm{x}|}$, with $x_0$ being the impact parameter, $L_{\mathrm{eff}}$ the effective interaction length, and $k_\mathrm{x}$ the PQP (imaginary) wavevector in the x-direction (further details in Supplementary Note 5).

The quantum theory of free-electron interaction with photons [72–75] shown in Eq. 1 can be used to show that the PQP emission should follow Poisson statistics (Supplementary Note 3). The Poisson distribution also appears in different ways in classical processes to reflect the mean distance (or time) between collision events. In our case, the Poisson parameter satisfies $\lambda = |g_{\mathrm{Qu}}|^2$, representing the average amount of emitted PQP quanta and indicating the quantum interaction strength. Based on this theory, we constructed a model to describe the distribution of energy loss in our system, combining the 2D-CR spectral density with the Poisson statistics:

$$\frac{dP}{du} = \underbrace{s\left(p \cdot \sum_{n=0}^{\infty} e^{-\lambda} \frac{\lambda^n}{n!} f_n(u) + (1-p)f_0(u)\right)}_{\text{Measured EEL signal}} + (1-s)\left(p \cdot \sum_{n=0}^{\infty} e^{-\lambda} \frac{\lambda^n}{n!} f_n(u) + (1-p)f_0(u)\right) \tag{2}$$

where $\frac{dP}{du}$ is the probability density that describes the EELS following all emission events as a function of the lost energy $u$. The probability density to emit $n$ PQP quanta, $f_n(u)$, is constructed from the previous order by the recursive convolution $f_n(u) = f_{n-1}(u) * f_{\mathrm{PQP}}(u)$, with $f_{\mathrm{PQP}}(u)$ being the spectral density of the PQP (see Supplementary Note 1 for its derivation). $f_0(u)$ is the initial energy distribution of the electron (also called the zero-loss peak; ZLP), $s$ is the probability to detect an electron in our setup, and $p$ is the probability that an electron interacted with the sample and hence was subjected to the Poisson process. This model implies two fitting parameters: (1) The Poisson parameter $\lambda$, which depends on the exact electron–photon interaction parameters (electron velocity, impact parameter, length of interaction); (2) The product $s \cdot p$, which is related to the experimental settings and detection capabilities. Eq. 2 gives a good fit to the EELS measurements for the parameters obtained from our experiment (solid red line and blue dots in Fig. 4(a,b), respectively).

**EELS as a method to quantify electron–photon coupling strength**

Figure 5 summarizes multiple EELS measurements with varying impact parameters and interaction lengths, showing the quantum coupling strength $g_{\text{Qu}}$ in each case. The effective interaction length of each spectrum is extracted from fitting it to theory, showing a good agreement with the measured values of the maximal interaction length (see Supplementary Note 5). The values of $g_{\text{Qu}}$ range between 0.51 and 0.99, directly implying that higher-order processes, such as multi-photon emission, are not negligible in our CR experiment (Supplementary Note 4). These $g_{\text{Qu}}$ values are more than two orders of magnitude larger than in previous free-electron experiments [2,5,6,61,75–82] (see comparison in Supplementary Note 6). Furthermore, these results are consistent with an ab initio model that uses macroscopic quantum electrodynamics [42,71] to calculate the coupling strength (see Supplementary Note 3). This model also shows the qualitative scaling of $g_{\text{Qu}}$: linearly dependent on the square root of the interaction length and exponentially dependent on the impact parameter.

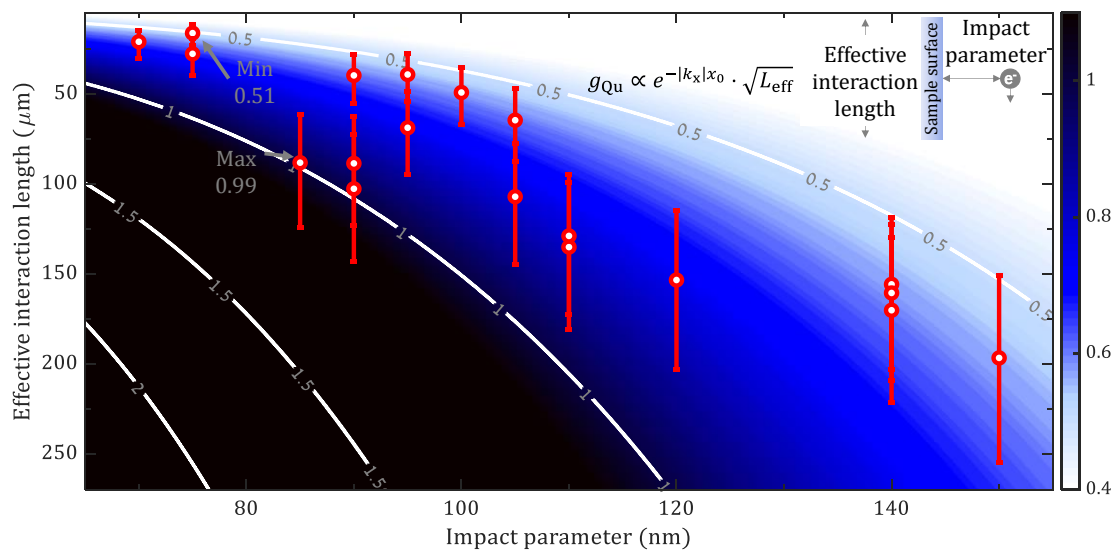

**Fig. 5 | Quantum coupling strength between free electrons and photonic quasiparticles.** The quantum coupling $g_{Qu} \propto e^{-|k_x|x_0} \cdot \sqrt{L_{eff}}$ values are extracted from 19 EELS measurements. Each value is overlaid on a theoretical calculation of the quantum coupling (colored background), derived in Supplementary Notes 1 and 5. The calculated $g_{Qu}$ is averaged over the electron beam Gaussian spatial distribution (with standard deviation of 30 nm; see Supplementary Note 5 for more details). The blue error bars represent a 10 nm difference in the impact parameter. The horizontal error bar for all data points is 10 nm in width, stemming from the fit resolution of the impact parameter (Fig. 3(c)). All the measurements in this figure are with 200 keV electrons.

**Discussion: does the electron become entangled with a single- or a multi-mode photonic state?**

The theory in Eq. 2 considers a single-mode description of the PQP, despite its bandwidth that may call for a multi-mode quantum-optical theory. The good match with the experimental results shows that the single-mode theory is accurate for predicting the EELS amplitudes. A full analysis of the 2D-CR effect should go beyond a single mode theory and consider the bandwidth of the emitted PQPs. Generally, the full quantum state of PQPs is composed of both (1) emission of multiple PQPs of different frequencies (e.g., $|1_{PQP,\omega_1}, 1_{PQP,\omega_2}, 1_{PQP,\omega_3}\rangle$), and (2) emission of Fock states of different orders (e.g., $|3_{PQP,\omega}\rangle$).

As recently described in a theory paper [69], determining whether the emission is a mixed state or a pure state depends on the (coherent) energy-uncertainty of the emitting electron. If the electron energy-uncertainty is smaller than the PQP energy bandwidth, then the electron emits distinguishable PQP states with different frequencies within the PQP bandwidth (case (1) above). Therefore, the resulting joint state contains entanglement between the electron and a multi-mode photonic state. In contrast, if the electron energy-uncertainty is much wider than the PQP energy bandwidth, then the electron emits pure PQPs, each containing the entire PQP bandwidth (case (2) above). Therefore, the resulting joint state contains entanglement only between the electron and multiple PQP Fock states of different orders. The coherence of each PQP Fock state in the latter case can be understood from noting that the electron cannot distinguish between different modes within the PQP bandwidth. Consequently, only in this latter case, the PQP may be rigorously considered as a single-mode photonic state.

To estimate the coherent energy-uncertainty of the electron, we can rely on a recent study that measured it to be around 0.3 eV at full-width half-maximum (FWHM) for the same electron source that we use, although in a different TEM system [83,84]. The bandwidth of the PQP in our experiment is less than 0.2 eV at FWHM. This supports the single-mode treatment in Eq. 2. An additional consideration altering the resulting photonic state of the PQP is its limited coherence length, which in our case is ~2μm due to optical losses. Since the coherence length is an order of magnitude shorter than the interaction length, the PQPs can be distinguished by their point of creation along the surface, creating a photonic state akin to case (1) instead of case (2), but distinguishable in space rather than frequency. Drawing a certain conclusion about the precise quantum state of the multi-PQP emission would require a more advanced quantum-optical detection, such as a cathodoluminescence scheme that involves autocorrelation measurements.

**Discussion: comparison with other multiple-peak EELS measurements**

It is insightful to compare our measurements to previous EELS studies that observed multiple loss peaks [48–56]. In all of those previous studies, the loss peaks emanated during the electron penetration through the sample or reflection from it, exciting matter resonances linked with non-propagating PQPs (e.g., bulk plasmons). In our case, the electron stays in vacuum and only interacts through the optical nearfield. We can now infer in hindsight, in light of the recent theoretical advances [67–70] and our analysis here, that those experiments [48–56] also excited a joint state of multiple PQPs entangled with the electron. The major difference is that all previous studies created such states with non-propagating PQPs, which are usually not considered for their photonic nature due to their short lifetime. In contrast, our work is the first to excite this behavior with propagating PQPs, which are often considered for their photonic nature since they may be coupled out of the sample [85].

Our findings critically depend on these properties of the propagating PQPs. In contrast to propagating PQPs, other matter excitations that are often measured with EELS [48–56], such as bulk and surface plasmons, cannot realize the Cherenkov effect because they are lossy, cannot propagate in the material, and occur at a single-frequency resonance independent of the electron energy. The distinction between the propagating PQPs and the non-propagating matter excitations can be extracted from the dielectric constants of the materials composing the structure in which they propagate. While non-propagating matter excitations appear as a single-frequency resonance at which the dielectric constant is purely imaginary, the propagating PQPs are guided at the interface between materials for which the (real parts of the) dielectric constants are one positive and one negative over a range of frequencies. This is why it is possible to excite them over a range of frequencies and use them to realize phenomena such as CR.

In all cases, our analysis shows the ability of measuring the quantum photonic state of free-electron radiation through the EELS of the emitting electron. As long as the electron energy uncertainty is narrower than the single photon energy, such measurements contain indirect information on both the diagonal and off-diagonal elements of the photonic density matrix, pointing on the electron-photon entanglement. In comparison, conventional photon counter detectors would have failed to identify the difference between the photonic state in our system and a Glauber coherent state, because they measure only the diagonal elements of the photonic density matrix [86], whereas the difference is hidden in the off-diagonal elements (Fig. 4(c) right column). Nevertheless, more work must be done to provide direct evidence of entanglement between the electron and the PQPs it emits. This for example could be done by coincidence measurement between the EELS and a cathodoluminescence detector capturing the emitted PQPs.

**Discussion: historical context for the quantum nature of free-electron radiation**

The history of research on the quantum nature of free-electron radiation, and particularly CR, goes back as far as 1940 [87–89]. Ginzburg [87] and Sokolov [88] were the first to describe CR using a quantum mechanical formalism – predicting changes that were considered negligible relative to the classical theory (given the experimental capabilities of that time). Recent theoretical papers showed additional, non-negligible, corrections to the classical CR theory that arise from the quantum nature of the electron – its wave properties, its spin, or its quantized orbital angular momentum [11,62,64]. The effect of the electron wavefunction in CR was even observed recently, through an inverse-CR experiment [61], building on contemporary demonstrations of quantum wavefunction-dependent features in *stimulated* processes of laser–electron interactions [2,75,90–95]. In contrast to the above-mentioned experiments, which exposed the quantum nature of electrons, our observation is based on a *spontaneous* process, which expose the quantum nature of photons.

For both spontaneous emission and stimulated emission, there are two necessary conditions for the quantum nature of photons to come into play. Only when both are satisfied, the emitted photons are entangled with the emitting electron. (1) The photonic state must be significantly modified, as can be quantified by the fidelity between the pre- and post-interaction photonic states. Stimulated interactions can usually be described classically because the initial photonic state is a Glauber coherent state that usually does not significantly change by emission or absorption (i.e., high fidelity). In contrast, in spontaneous interactions (as in our experiment), the initial photonic state is vacuum and thus emission of even just a single photon completely changes the photonic state (i.e., fidelity zero). (2) The electron state must be significantly modified, which can also be quantified by the fidelity between the pre- and post-interaction electron states. The fidelity may be estimated by the overlap between electron wavefunctions, as in Fig. 4(c) (noting that the relative phase must be considered as well). In our experiment, the electron energy uncertainty is narrower than the energy of the PQP, and thus the final electron becomes orthogonal (fidelity zero) to the initial electron. Since both conditions (1) and (2) are satisfied in our experiment, the emitted PQPs are in an entangled state with the emitting electron.

More generally, our results in 2D-CR suggest that analogous experiments should be able to reveal the inherent quantum features of other free-electron radiation phenomena. Such phenomena include transition radiation [18], Smith-Purcell radiation [96], 3D-CR [13,14], and other types of CR into confined PQPs in different structures. To reach electron–photon coupling strengths even stronger than what we observed, it is worth recalling that CR phenomena should appear in a wide variety of structure geometries, such as slab waveguides and van der Waals materials [97], or even lower dimensional structures, like slot waveguides [98] and metallic nanowires [57,99].

**Conclusion and outlook**

To conclude, we observed experimentally the free-electron emission of PQPs via the mechanism of 2D-CR, where the PQP dispersion relation caused the Cherenkov phase matching condition to occur in different frequencies for different electron energies. The dimensionality and bandwidth of 2D-CR allowed us to unveil the quantum photonic nature of the effect. The indirect evidence for electron-photon entanglement in our experiment could have intriguing consequences: For example, free electrons can provide a new way to generate quantum light such as single-photon [100] and multi-photon [101] Fock states, by post-selection on electrons with a certain energy loss [102]. Measurement of the electron after its light emission can also be used to infer what state it created without measuring (and destroying) the light state – thus providing a heralding mechanism for the emission of quantum light. Such approaches for providing indirect evidence of entanglement have only been shown before in very few cases in other kinds of interactions (e.g. [103,104]). For direct evidence of entanglement, one must employ coincidence detection between the electron and the emitted PQP, i.e., between the EELS and a cathodoluminescence detector [102].

Furthermore, our experiment delivered the strongest free-electron–light interaction to date, over two orders of magnitude stronger than previous experiments. Utilizing efficient coupling of the PQPs to free-space photons could help realize the prospects of electrons for bright quantum emitters, in line with predictions made over the past decade [10,100]. To show that the emitter is indeed very bright, we note that the effective lifetime of the 2D-CR process is few hundred femtoseconds (considering the interaction lengths, electron velocities, and extracted Poisson parameters in our experiments). This lifetime is consistent with the measured values of classical 3D-CR [28], but in a far smaller bandwidth. The combined short lifetime at a relatively narrow bandwidth makes the brightness of 2D-CR especially attractive.

Looking forward, we envision the use of 2D-CR and its inverse effect for integrated on-chip free-electron quantum emitters [8,9,100] and laser accelerators [6,76]. The interaction strength presented here is for the first time sufficient to allow single-electron–single-photon interactions, opening the door for free-electron cavity quantum electrodynamics [75,77]. This new interaction regime could enable the use of free electrons for quantum information applications [105,106] by entangling them with light [72], encoding them with qubit states [105], and utilizing them to entangle light in spatially-separated cavities [107].

* The first preliminary results of this study were presented in CLEO 2021 [108].

# APPENDIX A: EXPERIMENTAL METHODS

## 1. Sample preparation

A Si wafer was diced into 1×1 mm squares and subsequently sputtered (AJA International Inc. ATC 2200) with multiple layers in the following order: 5 nm Ti (adhesion layer), 200 nm Au, 10 nm $SiO_2$, and 30 nm $Si_3N_4$. We chose those layer thicknesses such that the energy of two-dimensional Cherenkov radiation (2D-CR) emission will be in the range of ~2–2.3 eV, to facilitate characterization in a transmission electron microscope (TEM). A square with an optical-grade surface was selected and then attached to a TEM sample holder with its diagonal parallel to the electron beam optical axis (see Fig. 2(e)).

## 2. Thickness measurement

Since the exact thickness of each layer greatly influences the PQP dispersion, the actual deposited film thicknesses were confirmed by cross-section imaging in another TEM (FEI Titan Themis G2). The lamella was prepared using a focused ion beam by a standard procedure (FEI Helios NanoLab DualBeam G3 UC). The layer thickness was measured to be 12.6 ± 1.2 nm for $SiO_2$ and 27.8 ± 1.1 nm for $Si_3N_4$ (average over 500×500 nm area). These numbers are used for the simulations in Fig. 3.

## 3. Electron energy loss spectroscopy (EELS)

EELS measurements were carried out in a Jeol-2100 Plus TEM. The system was designed to also operate as an ultrafast TEM, driven by femtosecond laser pulses, however, the current experiments did not use lasers. The electron beam was created by thermal heating of a $LaB_6$ tip and was accelerated to the kinetic energies specified in the main text. Using converged beam electron diffraction (CBED) mode, we created parallel electron illumination with about 30 nm beam diameter and aligned it to graze the sample surface along a length of up to 250 μm. The sample stage translation was used to control the impact parameter (beam-sample distance) and the estimated maximal interaction length (maximal path over the sample by the beam).

Electron energy loss spectroscopy (EELS) measurements were carried out using a Gatan GIF Quantum 965 spectrometer. Using low electron current, we configured the initial electron beam energy distribution (zero-loss-peak; ZLP) to be as narrow as 0.4-0.6 eV in its full-width half maximum (FWHM). The spectrally narrow electron beam facilitated the observation of individual peaks in the energy loss spectrum. The ZLP shape was recorded far from the sample and was later used in the data analysis process (See Supplementary Note 4). After the ZLP measurement, the EELS from the grazing interaction with the sample was recorded. The detector dispersion was ~0.01 eV per channel for all measurements. The combination of a narrow ZLP with high spectrometer dispersion allowed the capability of resolving the delicate shift in the 2D-CR emission energy.

## 4. Data analysis

To determine the exact peak position in Fig. 3(a) and 3(b), we carried out the following procedure: We normalized the EELS measurements and subtracted the ZLP from the EELS (this is a standard procedure that was for example also done in [75]). A single Gaussian function was matched to the first loss peak, and the mean of this Gaussian function was determined as the central frequency of the emitted PQP. To improve the sensitivity, we average the result over several independent measurements for each initial electron beam energy (this process was performed for each measurement separately). The final values in Fig. 3(c) are the averages over 2-12 repetitions of the same measurement in each energy. It is possible to resolve changes of the peak position with high accuracy (much better than the ZLP) due to the fact that the ZLP FWHM width was more than 4 times smaller than the PQP central energy (0.5 eV compared to ~2 eV), together with the high dispersion of our spectrometer.

We also provide in Fig. 3 the theoretical prediction as if the electrons in our experiment had interacted with the non-propagating surface plasmon resonance of our structure. The resonance frequency and width were determined using a full electromagnetic theory (Supplementary Note 1) that captures the interaction of an electron with the plasmonic resonance (the non-propagating PQP mode) without the interaction with the propagating PQP. This part of the energy loss spectrum does not depend on the electron energy, and thus we can simulate its Gaussian-style profile using electron energies under the Cherenkov threshold. We extract the specific height and width of the plasmon resonance EELS peak by simulating 30 keV electrons for the example in Figs. 3(a,b). The profile of the spectral shape (light green in Fig. 3(a)) is the result of a convolution with a 0.5 eV ZLP.

To extract the quantum coupling strength for each EELS measurement (as in Fig. 5 & Fig. 6), we carried out a nonlinear fitting process to find the correct theoretical description of those EELS curves. The full process is described in detail in Supplementary Note 4. We carried out the same fit procedure for different impact parameters to produce Fig. 3(c).

**APPENDIX B: PREDICTING THE PEAK FREQUENCY OF 2D-CR**

The most precise and rigorous way to obtain the spectrum of the 2D-CR and its peak emission frequency is by using the full electromagnetic description of the interaction as given in Supplementary Note 1. However, it is still worthy to describe a simpler theory that provides a more intuitive understanding of the phase-matching condition and determines the main trend of the peak emission frequency dependence on the electron velocity.

The Cherenkov phase-matching condition is mainly inferred from the intersection points on the below dispersion diagram between lines with slope equals to the electron energy and the curve of maximal imaginary part of the reflection coefficient, which is proportional to the density of states of the PQPs. These intersection points testify qualitatively on the experimental observation, because it predicts the red-shift of the peak emission frequency when decreasing the electron energy.

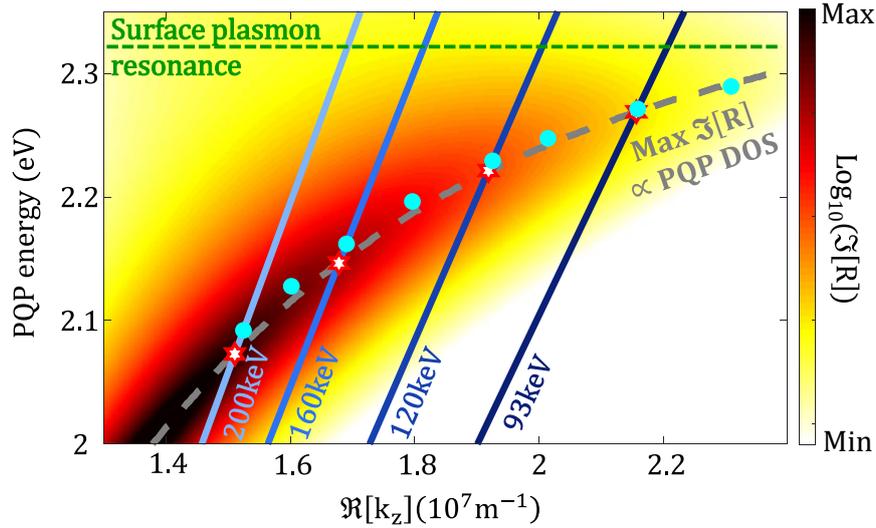

**Appendix Figure 1 | Deriving the Cherenkov phase-matching condition for co-propagating PQPs versus the full 2D-CR theory.** The colored background presents the imaginary part of our structure's reflection coefficient, which is proportional to the photonic density of states. The Cherenkov phase-matching condition can be approximately given by the intersection (red stars) between the maximal density of states (dashed gray curve) and lines with slopes that follow the electron velocity (shades of blue). The full electromagnetic calculation of the phase-matching condition (Supplementary Note 1) is also presented (cyan dots). The full calculation shows a slight blue-shift compared with the predictions made by the photonic density of states for most energies (i.e., the cyan dots are above the gray dashed line). The blue-shift arises from including emission into all angles, which is considered in the full calculation and not captured by the density of state consideration. For large wavevectors, the PQPs are more lossy, leading to another slight red-shift of the peak, which explains why for higher electron energies the full theory suddenly coincides with the PQP dispersion.

This is not the full picture though, since these intersection points reflect the phase-matching condition only between the electron and the PQP that is propagating parallel to it. Emission into PQPs that propagates in other angles can also satisfy the phase-matching condition and lead to a slight blue-shift of the peak frequency relative to the one expected from the reflection coefficient calculation. The full theory prediction is than back red-shifted for large wavevectors due to the lossy nature of the PQP there. In the main text Fig. 3(b) we used the full 2D-CR theory to compare with the experimental results.

**Acknowledgments**

The authors would like to thank Dr. Guy Ankonina for performing the evaporation process during sample preparation. With regard to the measurement of the layer widths, we thank Dr. Larisa Popilevsky, who prepared the TEM-lamella sample, and Dr. Galit Atiya, who acquired the EDX measurement. Samples were prepared at the Technion's Micro & Nano Fabrication Unit and their characterization performed at the electron microscopy center (MIKA) in the Department of Materials Science and Engineering at the Technion.

**Data availability**

The data supporting the findings of this study are available from the corresponding author upon reasonable request.